\documentclass[12pt]{article}
\usepackage{authblk}
\usepackage{bm}
\begin{document}

\noindent
{{\bf \Large {Monistic reduction of Einstein\rq{s} Equation for self-gravitating field masses}}}

%\subtitle{Monistic mechanics}
\medskip \noindent {I. {\'E}. Bulyzhenkov}

\noindent 
Levich Institute for Time Nature Explotations, %\orgaddress{\street{Street}, 
	{Moscow}, {119991}, Russia\\ 
	e-mail: ibphys@gmail.com, ORCID: 0000-0003-3835-0973

%\date{Received  March 2023 }
%\begin {abstract}
%\abstract
\bigskip \noindent{{\bf Abstract.} 
The divergence of relativistic accelerations generates field masses and the corresponding Ricci scalar in the gravitational integral of the Hilbert action. The covariant description of the elastic hierarchy with a specific time flow and Euclidean 3-section in the pseudo-Riemannian manifold is based on the monistic analogue of the Einstein Equation for a field mass with nonlocal inertia and self-gravity. The elastic autodynamics of metrically correlated densities correspond to Bianchi vector identities for material fields of continuous masses. The local time invariant  generates a primary reason for scalar mass densities and their local self-acceleration in a nonlocal hierarchy, rather than distant gravitational forces in the dualistic theory of pairwise interactions. Time should be studied experimentally as the (yin) inhomogeneous substance behind the (yang) observable densities of monistic matterspace with elastic hierarchies of nonlocal masses and their slow inelastic exchange.    
 }
%\end{abstract}

\bigskip%\keywords
\noindent
{\bf Keywords}: 
{	Spacetime hierarchies, elastic holism, monistic physics, autodynamics, Euclidean matterspace, invariant time rate  }
%\end {abstract}
%\maketitle

%\bigskip%\keywords
%\noindent  {\bf MSC 2020:} {83C25, 83C10, 83C22, 83A05, 83D05}

%\onecolumn {

%{\em E-mail: ibphys@gmail.com} %\\ ORCID: 0000-0003-3835-0973}}

\bigskip

\section {Duality in Einstein\rq{s} Equations}

		Logical iterations to find the relation between the Ricci tensor and the matter tensor on the basis of the dualistic model of geometrised fields and non-geometrised masses finally led Einstein to the tensor field equation on 2 December 1915
		 \cite{Ein1915}. 		
		 Today\rq{s} textbooks of such dualistic physics with pairwise fundamental interactions recommend deriving this famous equation of General Relativity  mathematically by varying the action of the immaterial field $\delta S_g\! = - (c^3/16\pi G)\delta\! \int\! R{\sqrt {-g}}d^4x$		 
		  $= (c^3/16\pi G)\!\int\! {\sqrt {-g}}d^4x [(g_{\mu \nu} R/2 - R_{\mu\nu} )\delta^{\mu\nu}$ $ - g^{\mu\nu}\delta R_{\mu\nu} ]$ 		
		and the action of matter $\delta S_m = \int T_{\mu\nu\nu}\delta g^{\mu\nu}{\sqrt {-g}}d^4x /2c $ by the metric tensor components $\delta g^{\mu\nu} \equiv \delta g^{\nu\mu}$ according to Hilbert\rq{s} technique \cite {Hil1915}, 
 \begin{equation}
 	 \frac {c^4}{8\pi G} \left ( g_{\mu\nu} \frac {R}{2} - 	R_{\mu\nu} \right) + T_{\mu\nu} = 0.
 \end{equation} 

	As a formal criticism of this variational equation, we note its mathematical incompleteness for non-stationary field states. In the most general case, local zeroing of the sign-variable subintegral densities is inadmissible, $g^{\mu\nu}\delta R_{\mu\nu}\neq 0$, although the fact that the corresponding integral $\int {\sqrt {-g}}d^4x g^{\mu\nu}\delta R_{\mu\nu} = \int d^4 x \partial_\mu ({\sqrt {-g}} w^\mu)$ = 0  can be nullified on the remote Gaussian surface \cite{LL}.
	 Einstein\rq{s} attempts to complete his field tensor $G_{\mu\nu} \equiv R_{\mu\nu} -g_{\mu\nu} (R/2)$ by a lambda term $g_{\mu\nu}\Lambda$ with a finite scalar density $\Lambda$ were not successful. Nevertheless, generalisations ${\tilde G}_{\mu\nu} = G_{\mu\nu} + \Lambda_{\mu\nu} $ for the Einstein tensor in equation (1) could contain such traceless contributions $\Lambda_{\mu\nu} $, which, under the tensor convolution $g^{\mu\nu}{\tilde G}_{\mu\nu} = g^{\mu\nu}{ G}_{\mu\nu} = -R $, would preserve the Ricci scalar density $R$ and the Bianchi geometric identities, $0 \equiv \nabla_\nu G^{\nu}_\mu = \nabla_\nu {\tilde G}^{\nu}_\mu$.

  Subsequent attempts by Einstein to improve on equation (1) were associ\-a\-ted with the negation of the tensor $T_{\mu\nu}$ from the dualistic world organisation. Einstein repeatedly called for a transition to a non-dual physics of material fields on the geometrical basis of the tensor $G_{\mu\nu}$ alone. Below we will continue these reasonable attempts by relating the scalar Ricci density $R>0$ to the local appearance of field masses in the monistic theory of continuous matter-space.

\section {Scalar sub-intervals in 3+1 interval}

Pseudo-Riemannian geometry for the 3+1 description of elastic hierarchies is very attractive because it can separate temporal and spatial sub-intervals in the scalar 4-interval $ds$ under general covariant transformations of coordinates in the hierachical reference frame,
\begin {equation}
ds^2 \equiv g_{\mu\nu} dx^\mu dx^\nu \equiv (g_\mu dx^\mu)^2 - (g_\mu  g_\nu - g_{\mu\nu})dx^\mu dx^\nu \equiv c^2 d\tau^2 - d^2 l.   
\end {equation}

The vectorial 4-potential $g_\mu \equiv g_{o\mu}/ {\sqrt {g_{oo}}}$ first appeared in relativistic physics when Einstein derived the geodesic law of motion of the point probe mass in the geometrised fields \cite {Ein1914}. 
    The geometric scalar $g_\mu dx^\mu/c \equiv d\tau$ for the local rate of physical time $\tau$ in a quasi-isolated hierarchy of elastic mechanical densities depends on all four coordinates of its specific spacetime. The difference of two geometric scalars $c^2d^2\tau - ds^2 \equiv (g_\mu g_\nu - g_{\mu\nu})dx^\mu dx^\nu \equiv (g_i g_j - g_{ij})dx^i dx^j \equiv d^2 l $ corresponds to such a relativistic scalar which, under general covariant transformations, does not go beyond the spatial 3-section. Here the metric tensor of the field 3-space $\kappa_{ij} \equiv g_ig_j - g_{ij}$ should not depend on the time parameter. Since physical time in the general case depends on changes of all four coordinates $x^\mu$, the metric subtensor $\kappa_{ij}$ for the 3-space section of a pseudo-Riemannian manifold should not depend on the spatial coordinates either,  
  $ g_ig_j - g_{ij}= \delta_{ij}  = const$ and $
   g^{ij}= - \delta^{ij}  =  const$.
   
    The Euclideanity of the common 3-space of overlapping elastic hirarhies (with the slow inelastic exchange for the undivided world evolution) must be a fundamental property for all spacetime organisations with  specific time rates $d\tau$ in the hierarical four-potential $g_\mu (x)\!\equiv\! \{g_{oo}(x)/g_o(x); g_{oi}(x)/g_o(x)\}$. Relatively weak inelastic exchange between quasi-elastic hierarchies of extended masses does not change the Euclidean geometry of their material spaces and this universal geometry for the common space of volumetric superpositions.
    
   \section {Holistic self-actions rather than interactions}
  
 Correlated accelerations of mass densities in elastic self-organisations of gravi\-tating matter do not obey the Newtonian scheme of pair interactions in the inelastic scenario of one-way energy transfer to the accelerated body under the action of external forces. The classical mechanics of inelastic forces in the laboratory has been very successful for the studied motion of the mass integral $m$, due to the fact that in practice the energy of push exchanges or external pulls is always much smaller than the elastic rest energy,  $mc^2  \gg \delta (mv^2/2)$. The \lq logical\rq{} application of the local push or pull of inelastic exchange from laboratory physics to distant external forces between point centres of elastic self-organisation has replaced the holistic self-action of continuously distributed mass and energy with the dual scenario of pair interactions in the void. This ontological trap has led to the dogma of the Standard Model of physical interactions based on four fundamental forces, which can hardly explain the elastic self-organisation of material densities with non-local feedback. 
 
 In holistic physics, there are no partners in the undivided whole, and correspondingly there are no interactions between non-existent partners. The\-re are only local self-actions for the correlated densities and stresses in each non-local hierarchy.  
 In reality, inelastic energy exchanges with corresponding force characteristics for elastic self-organisation of the solar system, for example, are so small that they have practically no influence on the laws of forceless self-acceleration of geodesically moving masses. 
 
 A quasi-elastic body can be successfully modelled to describe low-energy inelastic exchange through a point centre of mass, albeit with a huge rest energy. But it is unacceptable to assume a point body in empty space for the coherent description of internal geodesics within the continuous distribution of elastic densities of the extended body. On the contrary, every elastic hierarchy fills all space according to the laws of quantum mechanics. And spatial superpositions of densities of such extended bodies or continuous particles cross out the concept of empty space from the relativistic mechanics of elastic self-organisation with rest energy $mc^2$.
 
 The monistic physics of field-matter in extended quasi-elastic hierarchies with mutual superpositions and small inelastic exchanges can be based on the ponderable ether theory or the continuous hologram of de Broiglie material waves with their partial decoherence for slow world evolution. By neglecting inelastic exchanges for most cosmological applications, each cosmic hierarchy obeys the elastic holism for its field-mass densities with a constant integral $M = const$.  
  The holistic self-organisation of pure field matter can be described   
   monistically by the Hilbert action integral with the Ricci density $R > 0$.  
   This positive density corresponds to the scalar sum of the active and passive mass densities,   
  $ c^2 R/8\pi G = \mu_a + \mu_p = 2\mu_a = 2\mu_p$ \cite {Bul2008}. There is no need for additional energy-stress densities $T_{\mu\nu}$ in the monistic physics of filed masses.
 
{\section {Geometric densities of field masses and their elastic autodynamics}

The 2008 discovery of the relation of the Ricci scalar to static fields of radial mass or corresponding relativistic accelerations $a_i \equiv u^\nu\nabla_\nu u_i = - u^o \Gamma_{oi}^o u_o$ $= -\partial_i ln g_o$ now allows to specify $\nabla_\nu a^\nu = R/2$ in the Hilbert field action $- c^3/16\pi G \int d^4 x {\sqrt {-g}} R$. Six out of ten Hilbert variations go to zero for Eucli\-dean 3-space due to the constancy of its metric subtensor, $\delta g^{ij} \equiv \delta (-\delta^{ij}) = 0$. The remaining 4 degrees of geometrical freedom lead to the monistic analogue of Einstein\rq{s} Equation after variation of the field action by four independent components $g^{o\mu}$ of the metric tensor:
\begin{equation}
\frac {c^4}{8\pi G}\left ( g_{o\mu} \nabla_\lambda a^\lambda - \nabla_o a_\mu - \nabla_\mu a_o \right ) = 0 .
\end{equation}

 By rewriting this monistic field equation in the mixed components, we find $\delta^\nu_o \nabla_\lambda a^\lambda = \nabla_o a^\nu + \nabla^\nu a_o$, which gives the exact relations $\nabla_o a^o = \nabla_i a^i $, $\nabla_o a^i = -g^{i\nu}\nabla_\nu a_o = \delta^{ij}\nabla_j a_o -g^{io}\nabla_o a_o $. Based on the previous analysis of static mass densities \cite {Bul2008} one can introduce the tensor equivalence of passive (yin, time-time, $\mu_p\propto \nabla_o a^o$) and active (yang, space-space, $\mu_a\propto \nabla_i a^i$) mass densities in the time-varying elastic distributions with the scalar material density $R/2 =  \nabla_o a^o + \nabla_i a^i  =
  \partial_\nu a^\nu + a^\nu \partial_\nu ln g_o  $:
\begin{eqnarray}
	\cases { 
		%\mu_p + \mu_a = \frac {c^2}{8\pi G} R = 
	%	\frac {c^2}{4\pi G} ( \nabla_o a^o + \nabla_i %a^i),
	\frac {c^2\delta_o^\mu}{8\pi G}  \nabla_\nu a^\nu 
	=\frac {c^2}{8\pi G} (\nabla_o a^\mu + g^{\mu\nu} \nabla_\nu a_o)
	\cr
		\mu_p\equiv 
		\frac {c^2g^{o\nu}}{4\pi G} \nabla_\nu a_o =  \frac {c^2R}{16\pi G}
		= \frac {c^2}{4\pi G} \nabla_i a^i \equiv \mu_a,
		\cr 
	\frac {c^2\nabla_o a^i }{4\pi G}	= -\frac {c^2 g^{i\nu}}{4\pi G}\nabla_\nu a_o .
	%=	\frac {c^2}{4\pi G}(\delta^{ij}\nabla_j a_o %-g^{io}\nabla_oa_o).			
			} 
		\end{eqnarray}

  For static autodistributions of the field mass, when $g_{oi}=0, g_o = u_o = 1/u^o \neq 0,  u^i = u_i = 0 $, $a_o\equiv u^\nu \nabla_\nu u_o \equiv  u^\nu (\partial_\nu u_o - \partial_\nu u_o) = 0 $, the monistic field  equation (3) reads as $g_{oo}\nabla_\lambda a^\lambda = g^2_o R/2 = R_{oo} = 2\nabla_o a_o = - 2a_i \Gamma_{oo}^i = -2 (-\partial_i ln g_o) g_o \partial^i g_o $ or $ R = 2 (\nabla_o a^o + \nabla_i a^i) = \delta^{ij}\partial_i ln g_o \partial_j ln g_o$.
  The  equality of the passive (time-time, yin, unobservable) and active (space-space, yang, observable) masses or the scalar relation $\nabla^o a_o = \nabla_i a^i = \partial_i (-\delta^{ij} a_j)$ in (4) provides, for example,  the radial metric solution 
  $g_o = rc^2 / (rc^2 + G M)$ for spherically symmetric mass densities $\mu_a = \mu_p = GM^2c^2/ 4\pi r^2 (r c^2+ GM)^2$ \cite {Bul2008}. 
   
      Four time-varying balances $\nabla^o a_o = \nabla_i a^i $ and $\nabla_o a^i = \delta^{ij}\nabla_j a_o -g^{io}\nabla_o a_o$ in the monistic reduction (4) of ten Einstein\rq{s} equations make the self-gravitating dynamics of Euclidean matterspaces self-contained, since the relativistic metric solutions can be found independently of the Newton interaction limit or other gravitational theories.  Recall, that we do not re-postulate the Einstein Equivalance Erinciple for passive and active masses in the dualistic equation (1). Instead, we derive the equivalence of the time-varying mass densities from the monistic equation (3) after  raising its index $\mu$ (the fixed index $o$ cannot be raised).
      
      	 The last line in the relations (4)  is the local equivalence of the space-time and time-space countercurrents leading to  $g_i = \beta_i $ in equilibrium self-organisations (such as rotating galaxies). This three-vector equivalence for active and passive local currents 
      is essential for the self-contained theory to independently define three metric potentials $g_i \equiv g_{oi}/g_o \equiv g_{io}/g_o =\delta_{ij} g_o g^{oj} \equiv \delta_{ij} g_o g^{jo}$ for any non-equilibrium motions of self-assembling densities.
      Recall that the metric potential $g_o$ can be defined independently of Newtonian gravity from the local equivalence of the active and passive mass densities.

There is no need to add the stress-energy tensor $T_{\mu\nu}$ from point sources to the  monistic self-assemble of the inertial field with self-gravity.
Material singularities with distant interactions   led the metric theory of gravity  to the questionable  model of black holes and dark matter. With the  continuous filling of Euclidean space by massive fields without singularities, the new stress-energy tensor of monistic matter becomes the ten-component Hilbert density $T_{\mu\nu } \equiv c^4 B_{\mu\nu}/8\pi G \equiv c^4 (g_{\mu\nu} \nabla_\lambda a^\lambda - \nabla_\mu a_\nu - \nabla_\nu a_\mu)/8\pi G$  $\equiv c^4 g_{\nu\lambda} B^\lambda_\mu/8\pi G $ and 
$T^\mu_\mu  \equiv c^4 B/8\pi G \equiv c^4 \nabla_\lambda a^\lambda /4\pi G \equiv (\mu_p +\mu_a)c^2 $
according to the field relations in the variational theory \cite {Hil1915}. The vanishing divergence of the stres-energy tensor, $\nabla_\nu T^\nu_\mu = 0$, obeys the vector conservation law for elastic distributions of field matter: 
\begin{eqnarray}
0 =\frac {c^4}{8\pi G}\nabla_\nu B^\nu_\mu \equiv	\frac {c^4}{8\pi G}\left ( \nabla_\mu \nabla_\lambda a^\lambda - \nabla_\nu \nabla_\mu a^\nu  
	- \nabla_\nu \nabla^\nu a_\mu \right ) \cr \equiv 
		\frac {c^4}{8\pi G}\left (-R_{\mu\nu} a^\nu  
	- \nabla_\nu \nabla^\nu a_\mu \right ) = \frac {c^4}{8\pi G}\left ( \nabla_\mu \frac {R}{2} - \nabla_\nu R^\nu_\mu\right ).
\end{eqnarray}

Relativistic accelerations $a^\mu$  in the field self-organisation of elastic matter determine  both the local appearance of the mass density, $\nabla_\lambda a^\lambda = R/2 \propto \mu $, in equation (3), and the Lagrange-Hilbert autodynamics (5) under the scenario of Bianchi vector identities, where $\nabla_\nu (\nabla_\mu a^\nu 
+ \nabla^\nu a_\mu) = \nabla_\nu R^\nu_\mu =  \partial_\mu (\mu_p+\mu_a)4\pi G/ c^2$. The dynamical equation (5) has static solutions even for inhomogeneous mass distributions, when $\partial_i \mu_p = \partial_i \mu_a \neq 0. $ Thus the monistic audodynamics of inertial fields with self-gravity denies the Newnonian collapse of static mass densities. It is no longer surprising that the geodetic fall can elastically return probe masses from the relativistically dense region \cite {Bul2018}.The theoretical existance of static mass distributions in (5) and the elastic return \lq of the same\rq{} in the absence of inelastic exchange resolves the gravitational Bentley paradox for a  static hierarchy with metric fields.

\section {Discussion and Conclusion}

The monistic equation (3), unlike the dualistic equation (1), cannot be reduced to Newton\rq{s} theory with paired interactions-pulls through the void. On the contrary, the new tensor equations for the field auto-distribution of the mass integral with negligible inelastic exchange support the self-gravity of visible compactifications due to the local push \lq from here to there\rq{} of the Lomonosov etheric fluid \cite {Lom}. The Umov-Tsiolkovsky monism of the undivided world continuum with etheric mass-energies $kmc^2$, inertial heat, feedback adaptivity and nega-entropic self-assembly processes \cite {Umo, Tsi} is ignored by the black hole mainstream and the Stanard Model of paired interactions called fundamental forces.  There is  the lack of understanding of the conceptual role of the universal (Euclidean) geometry for the spatial overlap of all spatio-temporal self-organisations both in the quantum microcosm and in the macroscopic entanglement without inelastic exchange or dissipative measurements. 

Many science managers still do not understand the logic of scientific evidence \cite {Pop} and sincerely believe that the Schwarzschild curvature of space has finally been proven in practice without physical alternatives. But the field-mass metric with Euclidean space, $ds^2_{2008} = [rc^2 dx^o/(rc^2+ GM)]^2 - dr^2 - r^2 d\Omega^2$, is also successful in theoretically describing all the measured corrections to Newtonian gravity \cite {Bul2012}. The static metric of 1916 was logically criticised by Einstein because of unphysical singularities \cite {Ein1939}. There is no theoretical doubt about the scientific falsification of the Schwarzschild solution, and hence of the dualistic equation (1), as nonlocal experiments in space become more complicated and the accuracy of measurements increases.

When the inhomogeneous affine connections, which are related to the local stretching of physical time, go to zero in Minkowski spacetime, not only does the relativistic acceleration disappear, $a_i \Rightarrow - \Gamma_{oi}^o = 0$, but also the generation of local masses and field stresses in the Euclidean space of common observations stops. 
In this context, the origin of active and passive masses due to  self-actions in elastic spacetime hierarchies can be related to their primary course as physical time rate in metric autopotentials. On the basis of the temporal autogeneration of Euclidean field-mass densities in the monistic equation (3), one can support the conceptual conclusions of many Russian cosmists \cite {Ver, Chi, Koz} about the necessity to experimentally study time as a primary (unobservable, yin) energetic substance and the root cause of secondary (obsdrvable, yang) energetic events in the Euclidean matterspace of common observations.

\end{document}